\documentclass{aa}       
\usepackage{psfig}

\newcommand\lea{\mathrel{\raise .4ex\hbox{\rlap{$<$}\lower 1.2ex\hbox{$\sim$}}}}
\newcommand\gea{\mathrel{\raise .4ex\hbox{\rlap{$>$}\lower 1.2ex\hbox{$\sim$}}}}

\renewcommand\deg{\ifmmode^\circ\else$^\circ$\fi}

\def\cers#1.{{in: {\it Compact Extragalactic Radio Sources}, ed. J.A. Zensus
    and K.I. Kellermann (NRAO, Socorro), p.~#1.}}
\def\irvine#1.{{in: {\it Quasars and Active Galactic Nuclei: High-Resolution Radio Imaging}, ed.\ M.H. Cohen
    and K.I. Kellermann, {\it Proc.\ Nat.\ Acad.\ Sci.\ USA}, Vol.\ 92, No.\ 5, p.~#1.}}

\begin{document}

\thesaurus{(11.01.2; 11.10.1; 11.17.4:S5\,0836+710; 13.18.1) }

\title{VSOP imaging of S5~0836+710: a close-up on plasma instabilities in the jet}

\titlerunning{VSOP imaging of S5~0836+710}

\author{
	A.P.\ Lobanov\inst{1} \and
	T.P.\ Krichbaum\inst{1} \and
	A.\ Witzel\inst{1} \and
	A. \ Kraus\inst{1} \and
	J.A.\ Zensus\inst{1} \and
        S.\ Britzen\inst{1,2} \and
	K.\ Otterbein\inst{1,3} \and
	C.A.\ Hummel\inst{4}  \and
	K.\ Johnston\inst{4} \\
}

\authorrunning{A.P. Lobanov {\it et al.}}

\offprints{A.P. Lobanov}

\institute{Max-Planck-Institut f\"ur Radioastronomie, Auf dem H\"ugel 69, D-53121 Bonn, Germany
	\and
	The Netherlands Foundation for Research in Astronomy, P.O.Box 2, 7990 AA Dwingeloo, The Netherlands
	\and
	Landessternwarte Heidelberg, K\"onigstuhl, D-69117 Heidelberg, Germany
	\and
	United States Naval Observatory, Washington, D.C., USA
}

\date{Received 19 August 1998 / Accepted 21 September 1998}

\maketitle

\begin{abstract}

The luminous high-redshift ($z=2.17$) quasar S5~\object{0836+710} has
been observed at 5\,GHz with the VSOP. We compare the properties of
three images obtained from the observation: a low-resolution ground
array image (dynamic range 4600:1), a full-resolution VSOP image
(900:1), and an image made with only the space baselines (200:1). The
space baselines alone are sufficient for a reliable recovery of the
source structure, within the limits of the achieved spatial sampling
of the visibility data. The curved jet ridge line observed in the
images can be described by Kelvin-Helmholtz instabilities developing
in a relativistic outflow with the Mach number of about 6. This
description holds on the scales of $\lea 700\,h^{-1}$pc, and is shown to be
consistent with variable apparent speeds observed in the jet.


\end{abstract}

\section{Introduction \label{sc:intro}}

The VSOP (VLBI\footnote{Very Long Baseline Interferometry} Space
Observatory Program) is a Space VLBI (SVLBI) mission utilizing the
worldwide array of radio telescopes and an orbiting 8-meter antenna
deployed on the Japanese satellite HALCA\footnote{Highly Advanced
Laboratory for Communication and Astronomy} (\cite{hir96}).  The
satellite has an elliptical orbit, with the apogee at 21\,000\,km,
perigee at 560\,km, and orbiting period of about 6 hours. In each
observation, the data stream from the satellite is recorded by a
network of 5 STS (Satellite Tracking Station), and subsequently
correlated with the data from participating ground telescopes.
Regular VSOP observations started in September 1997 at 1.6 and 5\,GHz.
A 5\,GHz VSOP observation of \object{0836+710} (4C71.07, z$=$2.17) was
made on October 7, 1997. This radio source is an ultraluminous quasar
with well established, correlated (Otterbein et al. 1998) broad-band
variability in gamma-ray (\cite{fic94}), X-ray (\cite{bru94}), optical
(\cite{lin+93}), mm- and cm- radio regimes (Mar\-scher \& Bloom 1994).
The source has a well collimated VLBI-scale jet (Hum\-mel et al. 1992)
extending out to $\sim$180 milliarcseconds (mas). The jet is
substantially curved, with lateral displacements of its ridge line and
oscillations of its transverse width (possibly correlated with the
observed jet speeds,
\cite {kri+90}). The observed speeds change remarkably along the jet:
near the core, apparent speeds $\beta_{\rm app} \approx 10\,h^{-1}\,c$
are measured (with $q_0=0.5$ and $H_0 =
100h\,$km$\,$s$^{-1}$Mpc$^{-1}$); at 3$\,$mas core separation, the
speed decreases down to $\beta_{\rm app} \sim 2$-$3\,h^{-1}\,c$, and
then becomes larger once again, further out (Krichbaum et al. 1990,
Otterbein 1996). This kinematic behavior may be caused by MHD
instabilities developing in the jet plasma (Hum\-mel et al. 1992).

We present here a 0.2-mas (0.8$h^{-1}$\,pc) resolution VSOP image of
\object{0836+710}, and discuss the morphology and properties of the
compact jet in this quasar.

\section{VSOP observation and data reduction}

\begin{figure}
\centerline{\psfig{figure=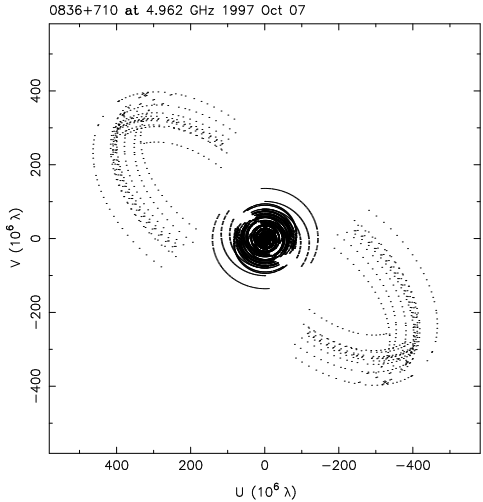,width=0.35\textwidth,angle=-90}}
 \caption[]{\label{fg:0836uvplot} The {\it uv}-coverage the VSOP
 observation of \object{0836+710}. The inner circles correspond to the
 baselines between ground radio telescopes; the space baselines are
 grouped into larger ellipses extending at a P.A.$\approx 45\deg$.}
\end{figure}

\object{0836+710} was observed with the VSOP at 5\,GHz for 11.5 hours,
with the VLBA\footnote{Very Long Baseline Array of the National Radio
Astronomy Observatory, USA} array (see Zensus, Diamond, \& Napier
1995, and references therein) providing ground support for the
observation. The data were recorded in the VLBA format, using total
observing bandwidth of 32\,MHz divided in two intermediate frequency
(IF) bands, each having 256 spectral channels.  The STS in Usuda
(Japan), Tidbinbilla (Australia), and Robledo (Spain) were used for
the satellite data acquisition. The data were correlated at the VLBA
correlator in Socorro, with output pre-averaging time of 1.966 and
0.524 seconds for the ground and space baselines respectively.  Fringe
visibilities were detected in the HALCA data recorded at Tidbinbilla
and Robledo. The resulting sampling function ({\it uv}-coverage) of
the final correlated dataset is shown in Figure
\ref{fg:0836uvplot}, indicating an improvement of resolution by a
factor of $\sim$3 compared to ground VLBI observations at the same
frequency.

\subsection {Fringe fitting}

\begin{figure}
\mbox{\psfig{figure=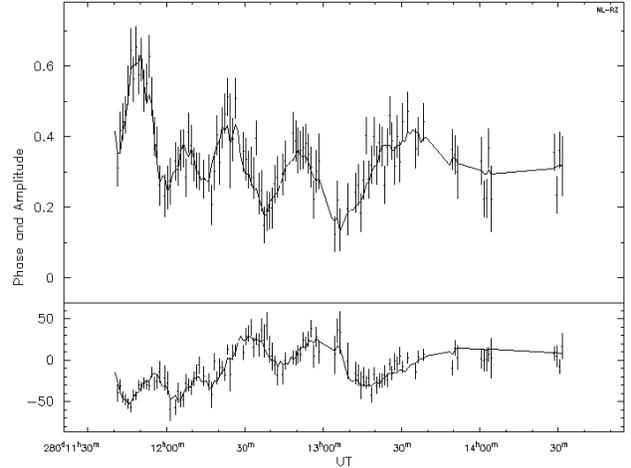,width=0.45\textwidth}}
 \caption[]{\label{fg:baseline} Visibility amplitudes (top panel) and
 phases (bottom panel) on the baseline between HALCA and VLBA:North
 Liberty, after amplitude calibration, fringe fitting, and phase
 self-calibration.  The lines drawn through the data represent the
 final CLEAN model of \object{0836+710}.}
\end{figure}

Post-processing of the correlated data has been done in
AIPS\footnote{The NRAO Astronomical Image Processing System} and
DIFMAP (\cite{she+94}).  We applied amplitude calibration, using the
antenna gain factors and system temperature measurements. After
inspecting the IF bandpasses, the last 46 channels were flagged in
each IF, owing to significant (50-80\%) amplitude reduction. This has
reduced the total observing bandwidth to 26.2 MHz. We applied the
phase-cal information available for the VLBA antennas, and then
corrected the residual delays and rates, using the single-band (SB)
and multi-band (MB) delay fringe fitting. We used solution intervals of 2
(SB) and 3 (MB) minutes, and accepted all solutions with ${\rm
SNR}>7$.  After the fringe fitting, the residual phase variations were
found to be within 3$\deg$ on the ground baselines, and $<10\deg$ on
the space baselines. We then averaged over all frequency channels, and
calibrated the phases with a point source model (to enable time
averaging). Finally, the data were exported into DIFMAP and further
time-averaged into 60-second bins. The amplitude and phase errors were
calculated from the scatter in the unaveraged data. Figure
\ref{fg:baseline} gives an example of typical HALCA baseline
visibility data. The estimated RMS noise on the HALCA baselines is
about 4.5 times higher than on the ground baselines.

\subsection{Imaging}

To investigate the impact of the VSOP mission, we have imaged both the
entire dataset (VSOP dataset hereafter) and only the ground baseline
data (VLBA dataset), and compared the properties of these images. Both
images have been produced after several cycles of hybrid mapping and
self-calibration.

\begin{figure*}[t]
\centerline{\psfig{figure=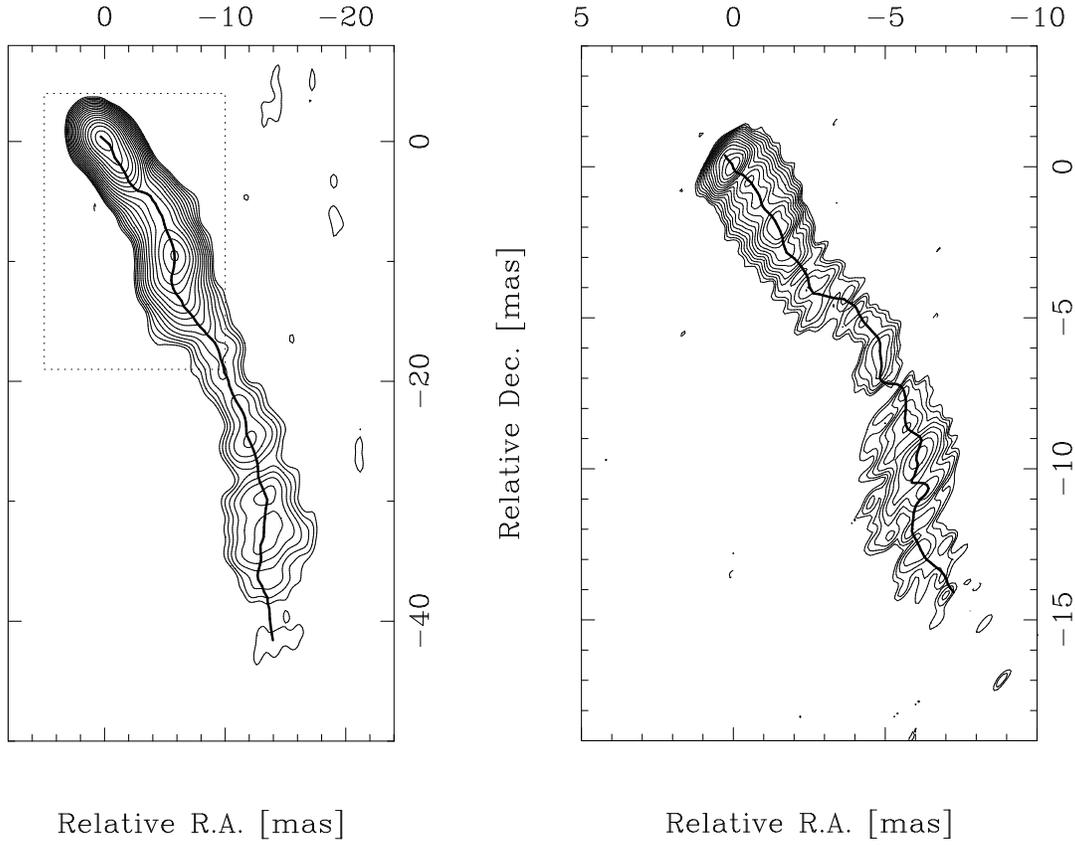,width=0.95\textwidth,angle=-90}}
\caption[]{\label{fg:0836maps} Ground array (left) and VSOP (right)
images of \object{0836+710}, with the jet ridge line marked. In the
ground array image, the dotted line rectangle shows the area covered
by the VSOP image. The image parameters are as follows. The ground
array image: restoring beam is 2.15$\times$1.75\,mas at
P.A.$=30.8\deg$; contour levels are drawn at $-$0.7, 0.7$\times
1.5^{n}$\,mJy/beam ($n=0,...,17$); the peak flux density is
918\,mJy/beam. The VSOP image: restoring beam is 0.93$\times$0.31\,mas
at P.A.$=-36.3\deg$; contour levels are drawn at $-$1.5, 1.5$\times
1.5^{n}$\,mJy/beam ($n=0,...,13$); the peak flux density is
432\,mJy/beam.}
\end{figure*}

We use natural weighting for gridding the
ground array data, enhancing the sensitivity to extended emission at
the price of slightly decreased image resolution. The gridding weights
are also scaled by amplitude errors raised to the power of $-1$. Both
phase and amplitude self-calibration have been used, with amplitudes
being allowed to vary only after the total model flux has approached
the zero spacing flux to within 3\%. The image obtained using this
procedure is shown in the left panel of Figure \ref{fg:0836maps}.

To image the VSOP dataset, we use uniform weighting which provides a
better angular resolution at the expense of lowering slightly the
sensitivity to extended structures. Because the estimated noise on
HALCA baselines is much higher than on the ground baselines, scaling
the gridding weights by the amplitude errors weights down
significantly all of the long {\it uv}-spacings. To avoid this, we
switch off the amplitude scaling of gridding weights. As a safeguard
measure, we apply only phase self-calibration to the VSOP
dataset. After a good fit to the data is achieved, we adjust the
antenna gains, correcting for small constant offsets between the model
and the visibility amplitudes. An example of the obtained fit to the
data can be found in Figure \ref{fg:baseline}. The final
high-resolution image is shown in the right panel of Figure
\ref{fg:0836maps}.

\begin{figure}
\centerline{\psfig{figure=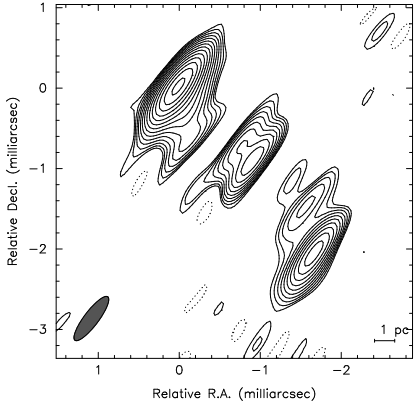,width=0.30\textwidth,angle=-90}}
 \caption[]{\label{fg:vonly} Image of 0836+710 obtained using only the
 baselines to HALCA. All structures seen in the image are consistent
 with those found in both the VSOP and VLBA images. This shows the
 reliability of the data recorded on the space baselines. The restoring
 beam is 0.68$\times$0.18\,mas at P.A.$=-36.9\deg$; contour levels are
 drawn at -3.2, 3.2$\times 1.5^{n}$\,mJy/beam ($n=0,...,11$); the peak
 flux density is 325\,mJy/beam. }
\end{figure}

To test the reliability of the satellite data, we make an additional
image using only the space baselines (HALCA image). The resulting
image shown in Figure~\ref{fg:vonly} is produced with only phase
self-calibration applied. The structures seen in Figure \ref{fg:vonly}
are consistent with the structures found in both VSOP and VLBA images
in Figure
\ref{fg:0836maps}. This reassures the good quality of the data
received from the orbiting antenna.

Table \ref{tb:images} compares the properties of all three
images. Note the roughly 5 times smaller dynamic range of the VSOP
image, which is caused by the poorer sampling and increased noise of
the data on the space baselines. The sufficiently high dynamic range
of the HALCA image (200:1) reflects once again the high quality of the
satellite data.

\begin{table}[h]
\caption{Parameters of the images \label{tb:images}}

\begin{center}
\begin{tabular}{l|ccccc}\hline\hline
Image & $S_{\rm tot}$ & $S_{\rm max}$ & $S_{\rm min}$ & $\sigma_{\rm RMS}$ &
$D^{\rm peak}_{\rm rms}$ \\
(1)  & (2) & (3) & (4) & (5) & (6) \\ \hline
VLBA & 2.167 & 0.918 & -0.001 & 0.2 & 4600:1 \\
VSOP & 2.156 & 0.432 & -0.003 & 0.5 & 900:1  \\
HALCA & 0.731& 0.325 & -0.007 & 1.4 & 200:1  \\ \hline
\end{tabular}
\end{center}
\medskip

Column designation: 2 -- total CLEAN flux [Jy]; 3 -- peak flux density
[Jy/beam]; 4 -- lowest negative flux density [Jy/beam]; 5 -- RMS noise
[mJy/beam]; 6 -- peak-to-RMS dynamic range.
\end{table}

\section{Properties of the compact jet}

\begin{figure}
\mbox{\psfig{figure=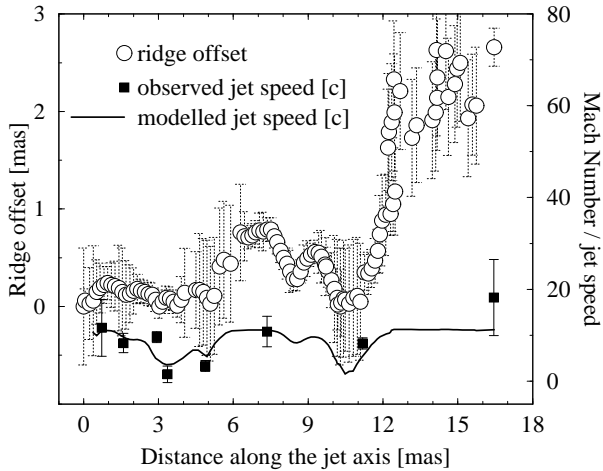,width=0.47\textwidth}}
 \caption[]{\label{fg:mach} Relation between the the ridge line and
 observed jet speeds in \object{0836+710}.  Circles are measured
 offsets of the jet ridge line from the jet axis at P.A.$=-146\deg$;
 squares denote speeds reported by Otterbein (1996) at different
 locations in the jet; solid line represents apparent speed variations
 derived from the measured ridge line offsets, for a jet with Mach
 number, $M_{\rm j}=6$, Lorentz factor, $\gamma_{\rm j}=11$, and
 opening half-angle, $\phi_{\rm j}=1\deg$.}
\end{figure}

In both VSOP and VLBA images, the jet is continuous, with several
enhanced emission regions embedded in it. In Figure \ref{fg:vonly},
curvature of the jet is noticeable even at $\lea 1$\,mas scales,
and it is consistent with the images obtained at higher frequencies,
most notably at 86\,GHz (Otterbein et al. 1998).  We consider now the
observed curvature of the jet represented, in the maps in
Figure~\ref{fg:0836maps}, by the curved ridge line of the jet.  We use
the median P.A.$=-146\deg$ to approximate the direction of the jet axis, and
measure the absolute values of corresponding offsets of the jet ridge
line. The offsets are plotted in Figure \ref{fg:mach}
(opaque circles).  One can see that the offsets show a remarkable
periodicity, possibly with several periods
superimposed. The derived offsets also appear to be correlated with
the apparent speeds measured along the jet (Otterbein 1996, black
squares in Figure \ref{fg:mach}).

The apparent curvature of the jet may reflect the presence of plasma
instabilities in the jet (e.g. MHD or Kelvin-Helmholtz instabilities).
The amplitude of ridge line displacements from the jet axis can be
used to estimate the jet Mach number, $M_{\rm j}$, using the
analytical approximation obtained by Hardee (1984) for the amplitude,
${\cal G}(r)$, of the fastest spatially growing mode of
Kelvin-Helmholtz instability: ${\cal G}(r) = {\cal G}(r_0)
(r/r_0)^{\epsilon}$. Here $r$ refers to the distance along the jet
axis, ``0'' denotes the initial values, and $\epsilon$ is determined
by the jet Mach number and opening half-angle, $\phi_{\rm j}$, so that
$\epsilon = 0.83/(M_{\rm j} \sin\phi_{\rm j})$.

To quantify the initial conditions of the jet plasma, we adopt the
results from Otterbein et al. (1998) obtained by modelling spectral
evolution in the core of the VLBI jet in 0836+710. The initial
amplitude of instability can then be approximated by the size of the
core, so that ${\cal G}(r_0) = 0.8$\,pc, at $r_0 = 15$\,pc
corresponding to the estimated distance from the jet nozzle to the
core.  With the reported jet bulk Lorentz factor, $\gamma_{\rm j}=11$,
viewing angle, $\theta_{\rm j}=3.2\deg$, and opening half-angle
$\phi_{\rm j}=1\deg$, we obtain the Mach number $M_{\rm j} \approx 6$,
within the innermost 3$\,$mas of the jet. It is similar to the value
reported in \cite{ott96}, and we take this as an argument in favor of
our choice of the direction of the jet axis. We expect the inner
jet to be pressure confined, with external pressure $P_{\rm
ext}\approx 25P_{\rm jet}$. Farther down the jet, at distances $>
10\,$mas ($> 700\,h^{-1}$pc), the derived values of $M_{\rm j}$ appear
to increase up to $\sim 30-50$, if we assume that the jet viewing
angle remains the same.  

Perhaps a more attractive possibility is to preserve the Mach number
along the jet, and instead use variations of the jet viewing angle for
explaining the observed ridge line offsets. For the same set of
$M_{\rm j}$, $\gamma_{\rm j}$, and $\phi_{\rm j}$, we determine the
variations of the viewing angle required to reproduce the offsets.
The derived $\theta_{\rm j}$ changes smoothly between $3\deg$ and
$65\deg$, with largest values occurring at $\sim 3$\,mas and $\sim
10$\,mas distances --- which are incidentally the locations where the
most pronounced lateral displacements are reported (\cite{kri+90},
\cite{ott96}).  The combination of the assumed $\gamma_{\rm j}$ and
derived $\theta_{\rm j}$ allows us to predict the apparent speed,
$\beta_{\rm app}$, along the jet.  The predicted $\beta_{\rm app}$
(plotted in Figure \ref{fg:mach} with the solid line) is in a
reasonable agreement with the measured jet speeds. At $r>13$\,mas, the
predicted $\beta_{\rm app}$ becomes flat, as it approaches the limit
of $(\gamma_{\rm j}^2-1)^{1/2}$. At distances $r>15$\,mas, a moderate
increase of $\gamma_{\rm j}$ up to $\sim 15$, could provide a better
agreement with the measured speed. However, at these distances, the
measurement uncertainties for both the speeds and ridge line offsets
are too large to warrant arguing in favor of variable bulk speeds along
the jet. We would like to emphasize that the achieved consistency between
the model and observed speeds is preserved, if the chosen
direction of the jet axis are within a few degrees from the one we
have used. In that case, only the derived jet Mach number would have
differed from $M_{\rm j}=6$ obtained above.  Sparsity of the available
speed measurements precludes better kinematic modelling and selecting
the jet axis unambiguously. We expect that making such a selection
would require determining the continuous velocity distribution along
the jet, for which a suitable monitoring program should be designed,
aimed at obtaining a sequence of high dynamic range and high fidelity
VLBI images of the source.

\section{Conclusions}

The reported observation of \object{0836+710} shows that the VSOP
mission provides an excellent opportunity for high-resolution and high
dynamic range imaging of VLBI-scale jets in extragalactic objects. The
obtained VSOP image of \object{0836+710} allows to investigate the jet
morphology and physical conditions on linear scales up to 1\,kpc. We
have studied the curved ridge line in the jet, and shown that, on the
scales of $\lea$700$h^{-1}$\,pc, it can be modelled by a pressure
confined relativistic outflow with Lorentz factor $\gamma_{\rm
j}\approx 11$, Mach number $M_{\rm j} \approx6$, and opening
half-angle $\phi_{\rm j} \approx 1\deg$. In this description, the two
main factors responsible for the observed curvature of the jet ridge
line are: 1)~Kelvin-Helmholtz instabilities developing in the
relativisitic plasma and 2)~variations of the angle, $\theta{\rm j}$,
between the velocity vector of the outflow and the line of sight. We
have shown that the derived variations of $\theta_{\rm j}$ can be used
for predicting the jet kinematics, and are consistent with the
observed variable apparent speeds along the jet.

\section*{Acknowledgements}

We gratefully acknowledge the VSOP Project, which is led by the
Japanese Institute of Space and Astronautical Science in cooperation
with many organizations and radio telescopes around the world.  We
thank R.~Porcas and J.~Marcaide for constructive comments on the
manuscript. The National Radio Astronomy Observatory is a facility of
the National Science Foundation operated under cooperative agreement
by Associated Universities, Inc.


\begin{thebibliography}{}

\bibitem[Brunner et al. 1994]{bru94} Brunner H., Lamer G., Staubert R., 
	 Worall D. M., 1994, A\&A 287, 436. 

\bibitem[Fichtel et al. 1994]{fic94} Fichtel C. E. Bertsch D. L.,  
	 Chiang J. et al., 1994, ApJS, 94, 551.

\bibitem[Hardee 1984]{har84} Hardee, P.E. 1984, ApJ, 287, 523.

\bibitem[Hirabayashi 1996]{hir96} Hirabayashi, H. 1996, Springer Science
        (Quarterly) 2, 11, 6.

\bibitem[Hummel et al. 1992]{hum+92} Hummel C. A., Muxlow T. W. B., 
	 Krichbaum T. P. et al., 1992, A\&A, 266, 93.

\bibitem[Krichbaum et al. 1990]{kri+90} Krichbaum T.P., Hummel C.A., 
	 Quirrenbach A. et al., 1990, A\&A, 230, 271.

\bibitem[von Linde et al. 1993]{lin+93} von Linde J., Borgeest U.,  
	 Schramm K.-J.  et al., 1993, A\&A 267, L23.

\bibitem[Marscher and Bloom 1994]{mb94} Marscher A. P. \& Bloom S. D.  
	 1994, \cers179.

\bibitem[Otterbein 1996]{ott96} Otterbein K., 1996, PhD thesis, University 
	 G\"ottingen.

\bibitem[Ot\-ter\-bein et al. 1998]{ott+98} Otterbein K., Krichbaum, T. P., Kraus, A., et al. 1998, 334, 489.


\bibitem[Shepherd et al. 1994]{she+94} Shepherd M. C., Pearson, T. J., \&
	Taylor, G. B. 1994, BAAS 26, 987.

\bibitem[Zensus, Diamond, \& Napier 1995]{zdn95} Zensus, J. A.,
	Diamond, P. J., \& Napier, P. J. (eds.) 1995, ``Very Long Baseline
	Interferometry and the VLBA'', ASP Conference Series, Vol. 82.

\end{thebibliography}
\end{document}